\documentstyle[fleqn,11pt,epsf]{article}

\catcode`\@=11
\long\def\@makefntext#1{
\protect\noindent \hbox to 3.2pt {\hskip-.9pt  
$^{{\eightrm\@thefnmark}}$\hfil}#1\hfill}		

\def\@makefnmark{\hbox to 0pt{$^{\@thefnmark}$\hss}}	
	
\def\ps@myheadings{\let\@mkboth\@gobbletwo
\def\@oddhead{\hbox{}
\rightmark\hfil\eightrm\thepage}   
\def\@oddfoot{}\def\@evenhead{\eightrm\thepage\hfil
\leftmark\hbox{}}\def\@evenfoot{}
\def\sectionmark##1{}\def\subsectionmark##1{}}

\newcounter{sectionc}\newcounter{subsectionc}\newcounter{subsubsectionc}
\renewcommand{\section}[1] {\vspace{12pt}\addtocounter{sectionc}{1} 
\setcounter{subsectionc}{0}\setcounter{subsubsectionc}{0}\noindent 
	{\tenbf\thesectionc. #1}\par\vspace{5pt}}
\renewcommand{\subsection}[1] {\vspace{12pt}\addtocounter{subsectionc}{1} 
	\setcounter{subsubsectionc}{0}\noindent 
	{\bf\thesectionc.\thesubsectionc. {\kern1pt \bfit #1}}\par\vspace{5pt}}
\renewcommand{\subsubsection}[1] {\vspace{12pt}\addtocounter{subsubsectionc}{1}
	\noindent{\tenrm\thesectionc.\thesubsectionc.\thesubsubsectionc.
	{\kern1pt \tenit #1}}\par\vspace{5pt}}
\newcommand{\nonumsection}[1] {\vspace{12pt}\noindent{\tenbf #1}
	\par\vspace{5pt}}

\newcommand{\textlineskip}{\baselineskip=13pt}

\def\abstracts#1#2#3{{
	\centering{\begin{minipage}{4.5in}\baselineskip=10pt\footnotesize
	\parindent=0pt #1\par 
	\parindent=15pt #2\par
	\parindent=15pt #3
	\end{minipage}}\par}} 

\renewenvironment{thebibliography}[1]
	{\frenchspacing
	 \ninerm\baselineskip=11pt
	 \begin{list}{\arabic{enumi}.}
        {\usecounter{enumi}\setlength{\parsep}{0pt}     
	 \setlength{\leftmargin 12.7pt}{\rightmargin 0pt} 
         \setlength{\itemsep}{0pt} \settowidth
	{\labelwidth}{#1.}\sloppy}}{\end{list}}

\newcounter{itemlistc}
\newcounter{romanlistc}
\newcounter{alphlistc}
\newcounter{arabiclistc}

\def\@citex[#1]#2{\if@filesw\immediate\write\@auxout
	{\string\citation{#2}}\fi
\def\@citea{}\@cite{\@for\@citeb:=#2\do
	{\@citea\def\@citea{,}\@ifundefined
	{b@\@citeb}{{\bf ?}\@warning
	{Citation `\@citeb' on page \thepage \space undefined}}
	{\csname b@\@citeb\endcsname}}}{#1}}

\newif\if@cghi
\def\cite{\@cghitrue\@ifnextchar [{\@tempswatrue
	\@citex}{\@tempswafalse\@citex[]}}
\def\citelow{\@cghifalse\@ifnextchar [{\@tempswatrue
	\@citex}{\@tempswafalse\@citex[]}}
\def\@cite#1#2{{$\null^{#1}$\if@tempswa\typeout
	{IJCGA warning: optional citation argument 
	ignored: `#2'} \fi}}

\def\@refcitex[#1]#2{\if@filesw\immediate\write\@auxout
	{\string\citation{#2}}\fi
\def\@citea{}\@refcite{\@for\@citeb:=#2\do
	{\@citea\def\@citea{, }\@ifundefined
	{b@\@citeb}{{\bf ?}\@warning
	{Citation `\@citeb' on page \thepage \space undefined}}
	\hbox{\csname b@\@citeb\endcsname}}}{#1}}

\def\@refcite#1#2{{#1\if@tempswa\typeout
        {IJCGA warning: optional citation argument
	ignored: `#2'} \fi}}

\def\refcite{\@ifnextchar[{\@tempswatrue
	\@refcitex}{\@tempswafalse\@refcitex[]}}

\def\pmb#1{\setbox0=\hbox{#1}
	\kern-.025em\copy0\kern-\wd0
	\kern.05em\copy0\kern-\wd0
	\kern-.025em\raise.0433em\box0}

\def\fnt#1#2{\footnotetext{\kern-.3em
	{$^{\mbox{\scriptsize #1}}$}{#2}}}

\font\tenrm=cmr10
\font\tenit=cmti10 
\font\tenbf=cmbx10
\font\bfit=cmbxti10 at 10pt
\font\ninerm=cmr9

\font\eightrm=cmr8

\def\qed{\hbox{${\vcenter{\vbox{			
   \hrule height 0.4pt\hbox{\vrule width 0.4pt height 6pt
   \kern5pt\vrule width 0.4pt}\hrule height 0.4pt}}}$}}

\begin{document}



\normalsize\textlineskip
\thispagestyle{empty}
\setcounter{page}{1}


\vspace*{0.88truein}

\centerline{NUOVO CIMENTO B 114, 1435-1440 (December 1999), [gr-qc/9910070 v3]}
\vspace*{0.065truein}
\centerline{ERMAKOV APPROACH FOR $Q=0$ EMPTY FRW MINISUPERSPACE OSCILLATORS}
\vspace*{0.035truein}
\vspace*{0.37truein}
\centerline{\footnotesize H. Rosu$^1$, P. Espinoza$^1$, and M. Reyes$^2$}
\vspace*{0.015truein}
\centerline{\footnotesize\it 1. Instituto de F\'{\i}sica,
Universidad de Guanajuato, Apdo Postal E-143, Le\'on, Gto, Mexico}
\baselineskip=10pt
\centerline{\footnotesize\it 2. Instituto de F\'{\i}sica y Matematica,
Universidad Michoacana, Apdo Postal 2-82, Morelia, Mich, Mexico}
\vspace*{10pt}
\vspace*{0.225truein}

\vspace*{0.21truein}
\abstracts{
{\bf Summary.} The Wheeler-DeWitt equation for empty
FRW minisuperspace universes of Hartle-Hawking factor
ordering parameter $Q=0$ is mapped onto the dynamics of a unit mass
classical oscillator.
The latter is studied by the classical Ermakov invariant method.
Angle quantities are presented in the same context.
}{}{}



\textlineskip                  
\vspace*{12pt}                 

\vspace*{1pt}\textlineskip	
\vspace*{-0.5pt}
\noindent


\noindent




\noindent






The formalism of Ermakov-type invariants [\refcite{erm}]
can be a useful, alternative method of investigating evolutionary and
chaotic dynamical problems in the ``quantum" cosmological
framework [\refcite{work}]. Moreover, the Ermakov method
is intimately related to geometrical angles and
phases [\refcite{book}], so that one may think of cosmological
Hannay's angles as well as various types of topological phases
as those of Berry and Pancharatnam [\refcite{dutta}].

Our purpose in the following is to apply the formal Ermakov scheme to the
simplest cosmological oscillators, namely the empty
Friedmann-Robertson-Walker (EFRW) ``quantum" universes.
When the Hartle-Hawking parameter for the factor ordering ambiguity is zero,
$Q=0$ [\refcite{hh}], the EFRW Wheeler-DeWitt (WDW)
minisuperspace equation reads [\refcite{rosu}]
\begin{equation} \label{1}
\frac{d^2\Psi}{d\Omega^2} -\kappa e^{-4\Omega}\Psi
(\Omega) =0~,
\end{equation}
where $\Omega$ is Misner's time related to the volume of space
$V$ at a given cosmological epoch through $\Omega=-\ln (V^{1/3})$ 
[\refcite{mi}],
$\kappa$ is the curvature index of the universe (1,0,-1 for closed, flat, 
open, respectively), and $\Psi$ is the wavefunction
of the universe. The general solution is obtained as a superposition of
modified Bessel functions of zero order, $\Psi(\Omega)=C_1I_{0}(\frac{1}{2}
e^{-2\Omega})+C_2K_{0}(\frac{1}{2}e^{-2\Omega})$ in the $\kappa =1$ case, and
ordinary Bessel functions  $\Psi(\Omega)=C_1J_{0}(\frac{1}{2}
e^{-2\Omega})+C_2Y_{0}(\frac{1}{2}e^{-2\Omega})$ in the $\kappa =-1$ case,  
where $C_1$ and $C_2$ are 
two arbitrary superposition constants (we shall work with $C_1=C_2=1$).
Eq.~(1) can be mapped 
onto the canonical equations for a classical point
particle of 
mass $M=1$, generalized coordinate $q=\Psi$,
momentum $p=\Psi ^{'}$,
evolving in Misner's time considered as Newtonian time for
which we shall keep the same notation.
Thus, one is led to
\begin{eqnarray}
\frac{dq}{d\Omega}&=&p~\\     
\frac{dp}{d\Omega}&=&\kappa e^{-4\Omega}q~.  
\end{eqnarray}
These equations describe the canonical
motion for a classical point EFRW universe as derived from the
time-dependent oscillator Hamiltonian of the inverted ($\kappa =1$)
and normal ($\kappa =-1$) type, respectively, [\refcite{inv}]
\begin{equation} \label{4}
H(\Omega)=\frac{p^2}{2}-\kappa e^{-4\Omega}\frac{q^2}{2}~.
\end{equation}
For this EFRW Hamiltonian the triplet of phase-space
functions $T_1=\frac{p^2}{2}$, $T_2=pq$,
and $T_3=\frac{q^2}{2}$ forms a dynamical Lie algebra (i.e.,
$H=\sum _{n}h_{n}(\Omega)T_{n}(p,q)$) which is closed with
respect to the Poisson bracket, or more exactly
$\{T_1,T_2\}=-2T_1$, $\{T_2,T_3\}=-2T_3$, $\{T_1,T_3\}=-T_2$. The EFRW
Hamiltonian can be written down in the form
\begin{equation} \label{H}
H=T_1-e^{4\Omega}T_3~.
\end{equation}
The Ermakov invariant $I$ belongs to the dynamical algebra
\begin{equation} \label{5}
I=\sum _{r}\epsilon _{r}(\Omega)T_{r}~,
\end{equation}
and by means of the invariance condition
\begin{equation} \label{6}
\frac{\partial I}{\partial \Omega}=-\{I,H\}~,
\end{equation}
one is led to the following equations for the unknown functions
$\epsilon _{r}(\Omega)$
\begin{equation} \label{7}
\dot{\epsilon} _{r}+\sum _{n}\Bigg[\sum _{m}C_{nm}^{r}h_{m}
(\Omega)\Bigg]\epsilon _{n}=0~,
\end{equation}
where $C_{nm}^{r}$ are the structure constants of the Lie algebra, that
have been already given above. Thus, we get
\begin{eqnarray} \nonumber
\dot{\epsilon} _1&=&-2\epsilon _2 \\
\dot{\epsilon} _2&=&-\kappa e^{-4\Omega}\epsilon _1-\epsilon _3\\
\dot{\epsilon} _3&=&-2\kappa e^{-4\Omega}\epsilon _2~.    \nonumber
\end{eqnarray}
The solution of this system can be readily obtained by setting
$\epsilon _1=\rho ^2$ giving $\epsilon _2=-\rho \dot{\rho}$ and
$\epsilon _3=\dot{\rho} ^2 +
\frac{1}{\rho ^2}$, where $\rho$ is the solution of the Milne-Pinney
equation [\refcite{mp}]
\begin{equation} \label{9}
\ddot{\rho}
-\kappa e^{-4\Omega}\rho=\frac
{1}{\rho ^3}~.
\end{equation}
In terms of the function $\rho (\Omega)$ and using (6),
the Ermakov invariant can be
written as follows [\refcite{l}]
\begin{equation} \label{10}
I=I_{{\rm kin}}+I_{{\rm pot}}=
\frac{(\rho p-\dot{\rho}q)^2}{2}+\frac{q^2}{2\rho ^2}=\frac{\rho ^4}{2}
\Big[\frac{d}{d\Omega}\left(\frac{\Psi}{\rho}\right)\Big]^2+\frac{1}{2}
\left(\frac{\Psi}{\rho}\right)^2.
\end{equation}
We have followed Pinney [\refcite{mp}] and Eliezer and Gray [\refcite{eg}] 
to calculate $\rho(\Omega)$ in terms of linear combinations of Bessel functions
such that the initial conditions given by these authors were fulfilled.
We worked with the set  $A=1$, $B=-1/W^2$ and $C=0$ of Pinney's constants, 
where $W$ is the Wronskian of the pair of Bessel functions. 
Moreover, we have chosen the angular momentum of the auxiliary Eliezer-Gray
two-dimensional motion as unity ($h=1$). Since $I=h^2/2$ we must get a 
constant half-unity value for the Ermakov invariant.
We have checked this by plotting $I(\Omega)$ for $\kappa=\pm1$ in 
Fig.~1.





In order to get angle variables, we calculate the time-dependent
generating function allowing one to
pass to new canonical variables for which $I$ is chosen as the
new ``momentum" [\refcite{l}]
\begin{equation} \label{11}
S(q,I,\vec{\epsilon}(\Omega))=\int ^{q}dq^{'}p(q^{'},I,\vec{\epsilon}
(\Omega))~,
\end{equation}
leading to
\begin{equation}  \label{13}
S(q,I,\vec{\epsilon}
(\Omega))=\frac{q^2}{2}\frac{\dot{\rho}}{\rho}+
I{\rm arcsin}\Bigg[\frac{q}{\sqrt{2I\rho ^2}}\Bigg]+
         \frac{q\sqrt{2I\rho ^2-q^2}}{2\rho ^2}~, 
\end{equation}
where we have put to zero the constant of integration.
Thus,
\begin{equation} \label{14}
\theta=\frac{\partial S}{\partial I}={\rm arcsin}
\Big(\frac{q}{\sqrt{2I\rho ^2}}\Big)~.
\end{equation}
Moreover, the canonical variables are now
\begin{equation} \label{15}
q=\rho \sqrt{2I}\sin \theta ~,\qquad
p=\frac{\sqrt{2I}}{\rho}\Big(\cos \theta+
\dot{\rho}\rho\sin \theta\Big)~.
\end{equation}
The dynamical angle will be
\begin{equation} \label{16}
\Delta \theta ^{d}=
\int _{\Omega _{0}}^{\Omega}
\langle\frac{\partial H_{\rm{new}}}{\partial I}\rangle
d\Omega ^{'}=
\int _{\Omega _{0}}^{\Omega}\Bigg[\frac{1}{\rho ^2}-\frac{\rho ^2}{2}
\frac{d}{d\Omega ^{'}}\Big(\frac{\dot{\rho}}{\rho}\Big)\Bigg]d\Omega ^{'}~,
\end{equation}
whereas the geometrical (generalized Hannay) angle reads
\begin{equation}  \label{17}
\Delta \theta ^{g}=\frac{1}{2}\int _{\Omega _0}^{\Omega}
\Bigg[\frac{d}{d\Omega ^{'}}
(\dot{\rho}\rho)-2\dot{\rho}^2\Bigg]
d\Omega ^{'}~.
\end{equation}
The sum of $\Delta \theta ^{d}$ and $\Delta \theta ^{g}$ is the
total change of angle (Lewis angle)
\begin{equation}  \label{16b}
\Delta \theta ^{t} =\int _{\Omega _{0}}^{\Omega}\frac{1}{\rho ^2}
d\Omega ^{'}~.
\end{equation}
Plots of the angle quantities (16-18) 
for $\kappa=1$ 
are presented in Figs.~2,3, and 4, respectively. Similar plots have been 
obtained for the $\kappa=-1$ case.












In conclusion, a cosmological application of the classical
Ermakov procedure has been presented here based on a
classical point particle representation of the EFRW WDW equation.
Finally, we notice that the Ermakov invariant
is equivalent to the Courant-Snyder one in accelerator physics
[\refcite{cs}]
allowing a beam physics analogy to
cosmological evolution.

\bigskip

\noindent
This work was partially supported by the CONACyT Project 458100-5-25844E.

\newpage

\nonumsection{References}

\newpage

\vskip 1ex
\centerline{
\epsfxsize=360pt
\epsfbox{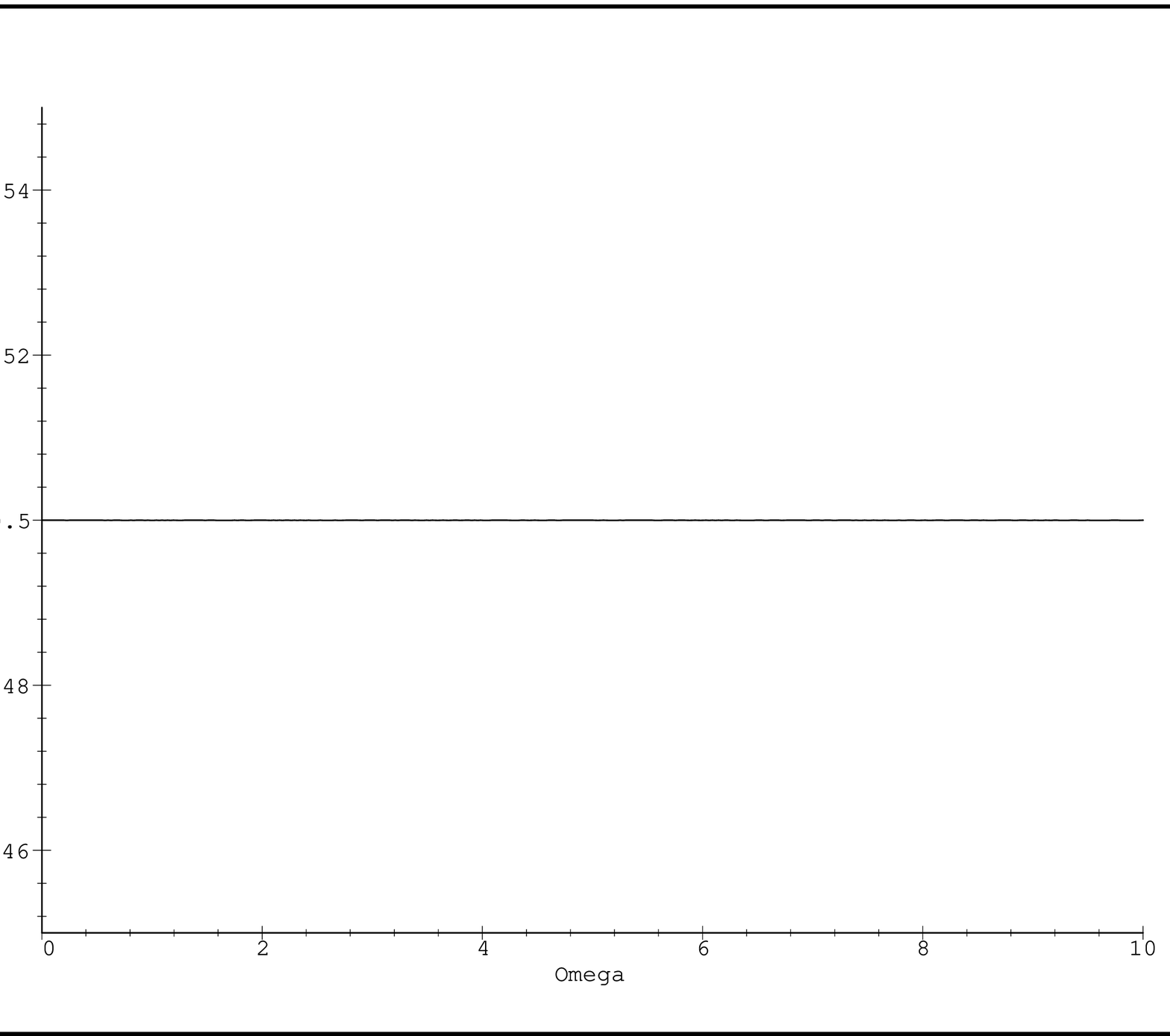}}
\vskip 2ex

\bigskip

\centerline{\footnotesize Fig.~1: The Ermakov invariant as a function 
of $\Omega$ for the closed EFRW minisuperspace model.}
\centerline{\footnotesize We got this plot for the open case as well.} 

\bigskip




\vskip 1ex
\centerline{
\epsfxsize=400pt
\epsfbox{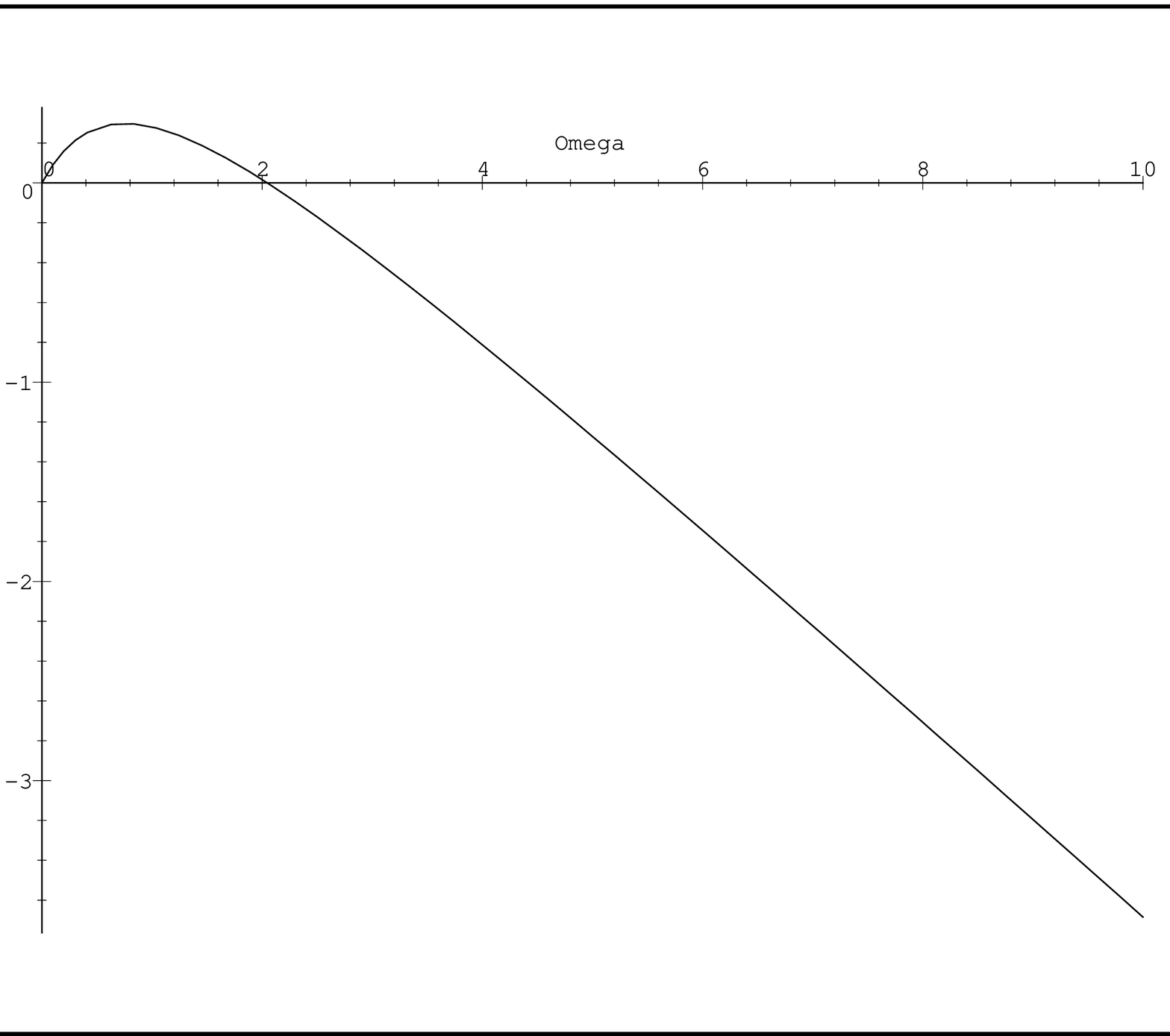}}
\vskip 2ex

\bigskip

\centerline{\footnotesize Fig.~2: The dynamical angle for the closed EFRW model.} 
%

\vskip 1ex
\centerline{
\epsfxsize=380pt
\epsfbox{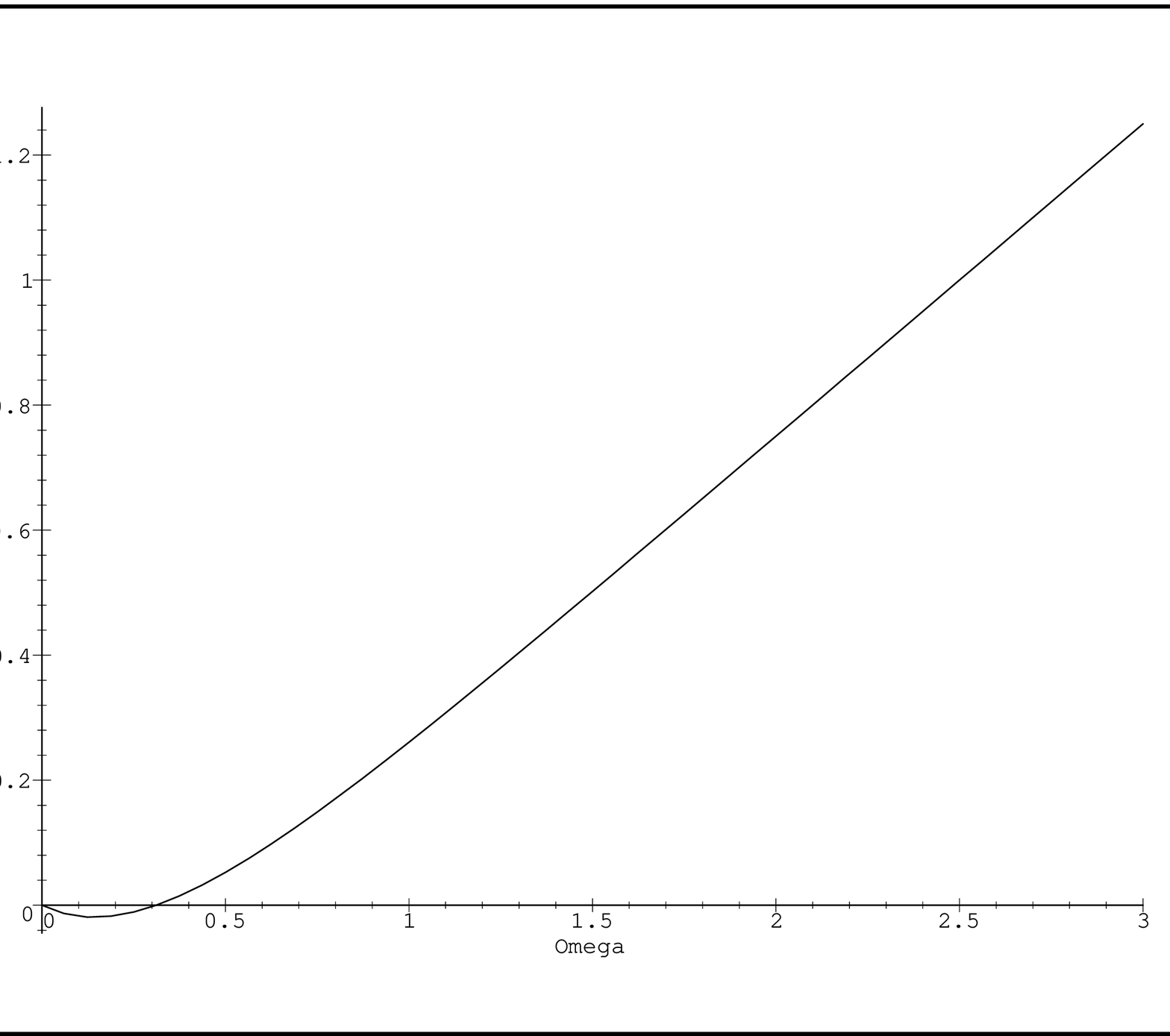}}
\vskip 2ex

\bigskip

\centerline{\footnotesize Fig.~3: The geometrical angle as a function 
of $\Omega$ for the same model.}

\vskip 1ex
\centerline{
\epsfxsize=380pt
\epsfbox{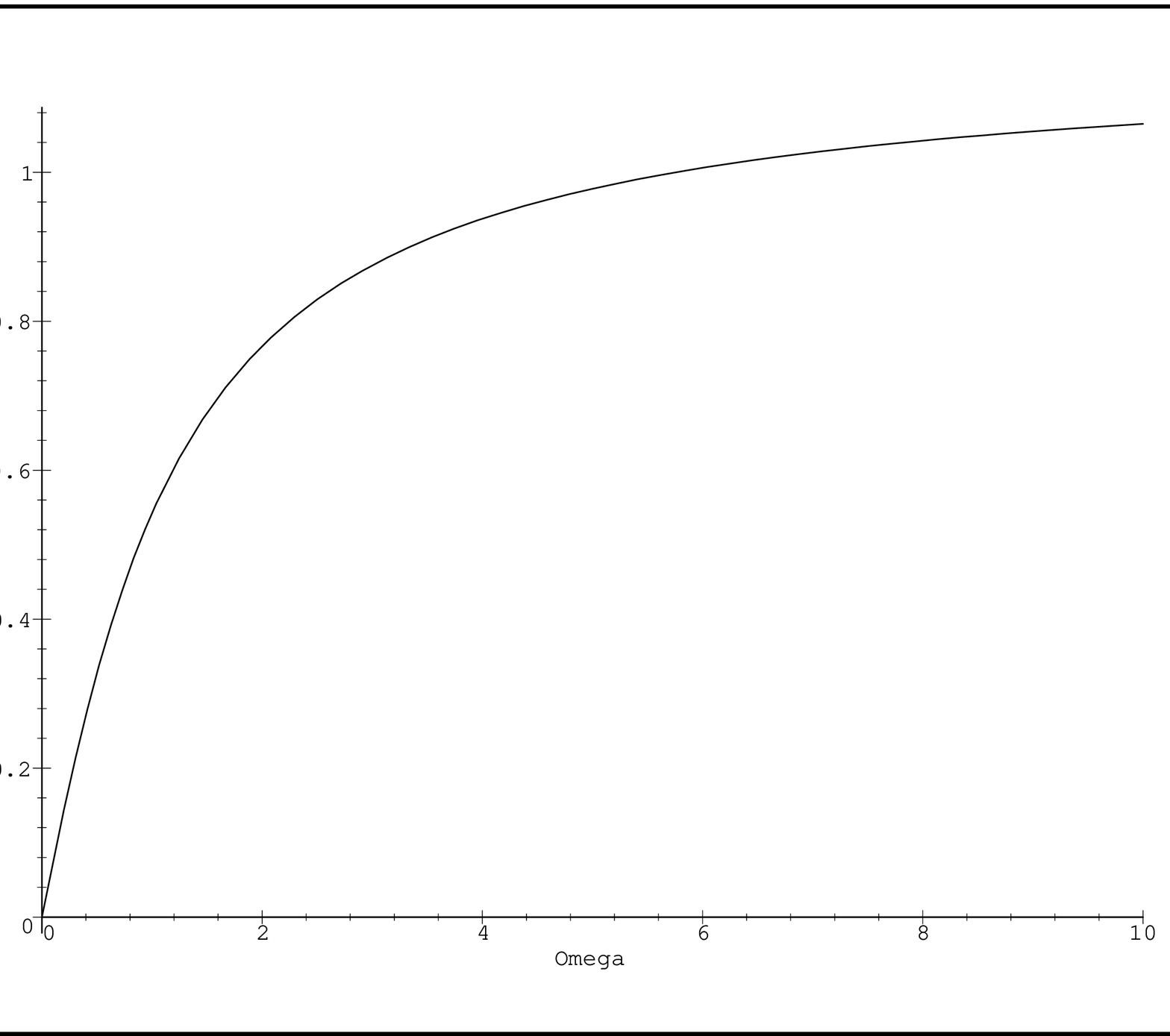}}
\vskip 2ex

\bigskip

\centerline{\footnotesize Fig.~4: The total angle as a function of $\Omega$ for 
the same model.} 

\bigskip

\end{document}

\end{document}